\documentclass[a4paper,11pt]{article}
\pdfoutput=1 

\usepackage{jcappub}

\usepackage{amsmath}
\usepackage{physics}

\def\beq{\begin{equation}}
\def\eeq{\end{equation}}
\def\bey{\begin{eqnarray}}
\def\eey{\end{eqnarray}}

\def\mpc{\, h^{-1}{\rm {Mpc}}}

\def\msun{\, h^{-1}{\rm {\rm M}_\odot}}

\def\gs{\mathrel{\raise1.16pt\hbox{$>$}\kern-7.0pt
\lower3.06pt\hbox{{$\scriptstyle \sim$}}}}
\def\ls{\mathrel{\raise1.16pt\hbox{$<$}\kern-7.0pt
\lower3.06pt\hbox{{$\scriptstyle \sim$}}}}
\def\gtsima{\, {\buildrel > \over \sim} \,}
\def\ltsima{\, {\buildrel < \over \sim} \,}
\def\prosima{\, {\buildrel \propto \over \sim} \,}
\def\gsim{\lower.5ex\hbox{\gtsima}}
\def\lsim{\lower.5ex\hbox{\ltsima}}
\def\simgt{\lower.5ex\hbox{\gtsima}}
\def\simlt{\lower.5ex\hbox{\ltsima}}
\def\simpr{\lower.5ex\hbox{\prosima}}

\def\log{{\rm log}}

\title{Halo formation and distribution around cosmic web}

\author[a,b]{Hongfei Yu}
\emailAdd{yuhongfei@mail.ustc.edu.cn}  

\author[a,b]{Huiyuan Wang}
\emailAdd{whywang@ustc.edu.cn}

\author[c,d]{Yangyao Chen}

\author[a,b]{Ziwen Zhang}

\author[a,b]{Hao Li}

\affiliation[a]{Department of Astronomy, University of Science and Technology of China, Hefei, Anhui 230026, China}
\affiliation[b]{
School of Astronomy and Space Science,
University of Science and Technology of China, Hefei 230026, China}
\affiliation[c]{School of Astronomy and Space Science, Nanjing University, Nanjing, Jiangsu 210093, China}
\affiliation[d]{Key Laboratory of Modern Astronomy and Astrophysics, Nanjing University, Ministry of Education, Nanjing, Jiangsu 210093, China}

\abstract{
We use the ELUCID simulation to study how the local environment affects the formation of dark matter halos and their spatial distribution around the cosmic web. We describe the environment by local density and cosmic-web type.
At fixed web type, the halo mass function (HMF) depends strongly on the local density, whereas at fixed density it depends only weakly, though significantly, on the web type; the character of both dependencies changes around the mean density. The halo formation redshift is primarily driven by local density rather than cosmic web. We then introduce the three-dimensional cosmic-web cross-correlation function (3D-WCCF), which expresses the distribution of halos around cosmic web in the local tidal eigenframe. The 3D-WCCF provides numerous anisotropic characteristics.
Around all web types, halos are most strongly correlated along the axis of weakest compression, $\mathbf{e}_3$, and least along the axis of strongest compression, $\mathbf{e}_1$. Nearly all cluster-sized halos reside in knots and lie preferentially in the plane perpendicular to ${\bf e}_1$. 
The 3D-WCCF also provides important information on the geometry and size of filaments, sheets and voids.
Moreover, early-forming halos are more strongly correlated with the web than late-forming ones, and this assembly bias is anisotropic around filaments, sheets, and voids, being strongest along $\mathbf{e}_3$, but nearly isotropic around knots.}

\begin{document}
\maketitle
\flushbottom

\section{Introduction}
\label{sec:intro}

Over the past decade, a growing body of work has examined how galaxies are related to the cosmic web, the network of knots, filaments, sheets, and voids into which matter is organized on large scales \citep{kraljic2018galaxy, alam2019cosmic, montero2024dependence, zhang2025unexpected, navdha2025relation, ullah2026cosmic}. Galaxy properties such as mass, color, and star formation activity vary with position in the web \citep{kraljic2018galaxy}, and dwarf galaxies show an unexpected clustering signal with respect to it \citep{zhang2025unexpected}. The cosmic web arises naturally from the gravitational growth of Gaussian initial fluctuations in a $\Lambda$CDM universe \citep{zel1970gravitational, white1987clusters, bond1996filaments}, and the dark matter within it collapses into virialized halos, inside which gas cools and galaxies form \citep{white1991galaxy, kauffmann1999clustering}. Thus, interpreting the galaxy--web relation requires a detailed understanding of how halos themselves are related to the web. This relation has two complementary aspects: how the properties of a halo depend on its environment, and how halos are distributed in space around the web.

Environments in which halos reside are commonly described either by the local density, smoothed on some scale, or the tidal force calculated from surrounding mass\citep[][]{wang2011internal}, or by a discrete cosmic-web type. The latter is often defined by the number of eigenvalues of the tidal tensor above a threshold \citep[T-web;][]{hahn2007properties, hahn2009tidal, foreroromero2009dynamical}; other schemes use the velocity-shear tensor \citep[V-web;][]{hoffman2012kinematic} or the density field directly \citep[NEXUS;][]{cautun2013nexus}. Large-scale structures are typically categorized into knots, filaments, sheets, and voids. In contrast to the local density parameter, cosmic web classification also captures anisotropy information about large-scale structures.
For a systematic comparison of these methods, see \citet{libeskind2018tracing}. 

The most basic property of halos is the halo abundance or the halo mass function (hereafter HMF). In the excursion-set framework the HMF follows from the statistics of the initial density field \citep{bond1991excursion}, and its conditional form predicts that dense regions host more massive halos \citep{bower1991evolution, lemson1999environmental}. Ellipsoidal-collapse models incorporate the dependence of collapse on the initial tidal shear and improve agreement with simulated halo abundances \citep{sheth2001ellipsoidal, sheth2002excursion}. However, they do not directly predict halo abundances conditioned on a specified cosmic-web classification.  
Simulations show that the HMF varies strongly with the web type \citep{hahn2007properties}. However, the local density and web type are not independent \citep{hahn2007properties, metuki2016abundance}. Whether the web type affects halo abundance beyond the density has been studied in different configurations which differ the conclusions. 
Using a spherical-collapse model together with an effective-universe approach, \citet{alonso2015halo} showed that the conditional mass function is determined solely by the environmental density. They also reported from simulations that, at fixed density, the HMFs across different web environments are consistent with one another to within $\sim13$ percent. In contrast, \citet{metuki2016abundance}, using a V-web classification, found that the mass function depends on the web type but not the ambient density. More research is certainly required.

Halo assembly depends also on the local density and cosmic web. Simulations show that halos in denser regions form slightly earlier \citep{sheth2004environment}, and that halo clustering depends on formation time, concentration, and spin \citep[e.g.][]{gao2005age, wechsler2006dependence, gao2007assembly, croton2007assembly}. This secondary, or assembly, bias matters both for interpreting halo clustering and for modeling how galaxies populate halos \citep{zentner2014galaxy}. \citet{hahn2007properties} reported that low-mass halos are older in knots and younger in voids. 
\citet{yang2017revealing} showed that assembly bias is stronger in knots and filaments than in sheets and voids. Assembly bias has also been linked to the proximity to cosmic-web features in simulations \citep{montero2024dependence} and detected environmental assembly bias in observations, with a stronger dependence on local density than on cosmic-web distances \citep{rodriguez2026connecting}.
Several mechanisms have been proposed to explain assembly bias. 
\citet{wang2007environmental} suggested that the tides from neighboring massive halos can suppress the growth of low-mass halos and contribute to age-dependent clustering\citep[see also][]{hahn2009tidal,wang2011internal}.
\citet{dalal2008halo} connected its high-mass behavior to the statistics of initial density peaks and argued that, at low masses, a substantial contribution arises from halos whose accretion has ceased. Correcting halo accretion histories for such arrested development reduces the assembly bias of low-mass halos by up to a factor of two \citep{smith2024reversing}, and backsplash subhalos also contribute a large part of the assembly bias\citep{wang2009distribution, mansfield2020three}. The tidal anisotropy of the local environment has been found to mediate the correlation between internal halo properties and large-scale bias over a wide range of masses \citep{paranjape2018halo, ramakrishnan2019cosmic}. 

The spatial distribution of halos and galaxies around the cosmic web carries information that the dependence on local environment does not. While the tidal anisotropy measures the intensity of shear, the halo distribution relative to the principal axes of the tidal field adds details on the directions. Working in the tidal frame of the saddle points of the potential, \citet{musso2018cosmic} predicted that halo mass, accretion rate, and formation time vary with orientation relative to a filament as well as with distance from it, with early-forming, mass-starved halos lying preferentially along the filament axis, and \citet{kraljic2018galaxy} interpreted the variation of galaxy properties with distance to filaments in the same framework. Building on the same framework, \citet{shim2026filaments} further showed that the anisotropic tidal environment most strongly influences halo formation on a scale of 2--3 times the Lagrangian radius of the halo, so that the relevant scale of the tidal field depends on halo mass. Directional measurements have so far used a frame attached to the halo: taking each halo's major axis as the reference direction, \citet{ma2026anisotropic} found the anisotropic assembly bias of halos to be strong for spin and shape but weak for formation time and concentration. The distribution of halos around the web itself, on the other hand, is usually quantified with isotropic statistics, such as the halo bias in different web environments \citep{yang2017revealing, balaguera2024secondary} or the cross-correlation between web environments and tracers \citep{zhang2025unexpected}, which average over the direction of each tracer relative to the local tidal field. A measurement that refers the halo distribution to the eigenframe of the web, and keeps all three directions, is still lacking.

In this paper, we use the ELUCID simulation \citep{wang2016elucid} to examine how halos form and how they are distributed in space within different cosmic-web environments, the latter measured with a cross-correlation function resolved along the local tidal eigendirections. This paper is organized as follows. Section~\ref{sec:simulation} describes the ELUCID simulation, the tidal-tensor web classification, and the halo samples. Section~\ref{sec:halo_env} separates the roles of the local density and the web class, using the halo mass function and the mean formation redshift. Section~\ref{sec:web_halo_correlation} introduces the three-dimensional cosmic-web cross-correlation function and presents the  anisotropy of the halo--web correlation, and examines its dependence on halo formation time. We summarize our results in Section~\ref{sec:summary}. 

\section{Simulation, Cosmic Web and Halos}
\label{sec:simulation}

\subsection{The ELUCID simulation}

In this work, we use the ELUCID (Exploring the Local Universe with reConstructed Initial Density field) simulation \citep{wang2016elucid}, a large dark-matter-only cosmological $N$-body simulation designed to reproduce the large-scale structure in the local Universe. The simulation was carried out using the \textsc{L-GADGET} code, a memory-optimized version of \textsc{GADGET-2} \citep{springel2005cosmological}. It follows the evolution of $3072^{3}$ dark matter particles in a periodic cubic box of side length $500\,h^{-1}{\rm Mpc}$. The corresponding particle mass is $3.08\times10^{8}\,h^{-1}{\rm M_{\odot}}$.
The simulation adopts a flat $\Lambda$CDM cosmology consistent with the WMAP5 results \citep{dunkley2009five}, with cosmological parameters $\Omega_{\rm m}=0.258$, $\Omega_{\Lambda}=0.742$, $\Omega_{\rm b}=0.044$, $h=0.72$, $\sigma_{8}=0.80$, $n_{\rm s}=0.96$.
The simulation outputs 100 snapshots from $z=18.4$ to $z=0$, equally spaced in the logarithm of the expansion factor.

\subsection{Halo catalog and halo formation time}\label{halo_catalog}

Dark matter halos are identified using the friends-of-friends (FoF) algorithm \citep{davis1985evolution} with a linking length of $b=0.2$ times the mean inter-particle separation. Gravitationally bound substructures within each FoF halo are further identified with the \textsc{SUBFIND} algorithm \citep{springel2001populating}. In each FoF system, the most massive subhalo is defined as the central or main halo, and its most-bound particle is used as the halo center. Halo masses ($M_{\rm h}$) are defined as the mass enclosed within a spherical radius $R_{\rm h}$ within which the mean density is 200 times the mean density of the Universe. In order to ensure reliable measurements of halo properties, we only use halos containing at least $N_{\rm p}=800$ dark matter particles, corresponding to a minimum halo mass of $M_{\rm min} \simeq 2.46\times10^{11}\,h^{-1}{\rm M_{\odot}}$.

In order to avoid complications associated with tidal stripping and satellite evolution inside larger host systems, our analysis is based on distinct host halos identified at $z=0$ in the ELUCID simulation. Therefore, we also exclude halos with a backsplash history, i.e., halos that were previously subhalos of a more massive host.

Halo merger trees are constructed by linking subhalos between adjacent snapshots. For a given subhalo at one snapshot, its descendant is identified as the subhalo in the subsequent snapshot that contains the largest weighted number of its member particles, where particles are weighted according to their binding energies. The main progenitor branch of each $z=0$ main halo is then used to trace its mass assembly history. For each halo,  from the main progenitor branch, we compute the halo formation redshift, $z_{\rm f}$, defined as the redshift at which the main progenitor first reaches half of the final halo mass at $z=0$.

\subsection{Cosmic web classification and local eigenframe}

To characterize the large-scale environment of halos, we construct the matter density field from the dark matter particles of the ELUCID simulation. The particles are assigned to a regular Cartesian grid of $500^{3}$ cells using the cloud-in-cell (CIC) scheme \citep{birdsall1969clouds} to construct the density field $\rho(\mathbf{x})$. The density contrast is defined as,
\begin{equation}
    \delta(\mathbf{x}) =(\rho(\mathbf{x})-\bar{\rho})/{\bar{\rho}},
\end{equation}
where $\bar{\rho}$ is the cosmic mean density. The density field is smoothed with a Gaussian kernel of scale $R_{\rm s}$. Hereafter $\delta$ denotes the smoothed density contrast.

The rescaled gravitational potential $\Phi$ is obtained from the overdensity field through the Poisson equation, $\nabla^{2}\Phi(\mathbf{x}) = \delta(\mathbf{x})$.  The tidal tensor is then defined as
\begin{equation}
    T_{ij}(\mathbf{x}) =\dfrac{\partial^{2}\Phi(\mathbf{x})}{\partial x_i \partial x_j}.
\end{equation}
At each grid cell, the tidal tensor is diagonalized to obtain three eigenvalues, $\lambda_1\geq\lambda_2\geq\lambda_3$, and the corresponding orthonormal eigenvectors, $\mathbf{e}_1$, $\mathbf{e}_2$, and $\mathbf{e}_3$. These eigenvalues are used to classify the cosmic-web environment following the T-web scheme \citep{hahn2007properties, hahn2009tidal}:
\begin{itemize}
    \item Knot: $\lambda_3 > \lambda_{\rm th}$, $\lambda_2 > \lambda_{\rm th}$, $\lambda_1 > \lambda_{\rm th}$
    \item Filament: $\lambda_3 < \lambda_{\rm th}$, $\lambda_2 > \lambda_{\rm th}$, $\lambda_1 > \lambda_{\rm th}$
    \item Sheet: $\lambda_3 < \lambda_{\rm th}$, $\lambda_2 < \lambda_{\rm th}$, $\lambda_1 > \lambda_{\rm th}$
    \item Void: $\lambda_3 < \lambda_{\rm th}$, $\lambda_2 < \lambda_{\rm th}$, $\lambda_1 < \lambda_{\rm th}$
    \label{web_classification}
\end{itemize}
while the eigenvectors define the local tidal eigenframe used in our anisotropic correlation analysis. 

For an overdensity $\delta$, the mass contained in the Gaussian kernel is 
\begin{equation}
   \log M_{\rm s}/\msun\equiv\log((2\pi)^{3/2}R_{\rm s}^{3}\bar{\rho}(1+\delta)). 
\end{equation}
Consider a halo of $M_{\rm h}$ placed in a cell with $\delta$. When $M_{\rm h}\ll M_{\rm s}$, the density of the cell is determined mainly by the surrounding environment; in contrast, if this is not the case, $\delta$ is largely determined by the halo itself. $M_{\rm s}$ can also be viewed as the maximum halo mass expected within a cell of overdensity $\delta$.
We adopt $R_{\rm s}=2\,h^{-1}{\rm Mpc}$ and $\lambda_{\rm th}=0$ as our fiducial configuration, a common choice in the literature\citep[see e.g.][]{hahn2007properties}. For $R_{\rm s}=2\mpc$, we have $\log M_{\rm s}=12.95+\log(1+\delta)$.

\begin{figure*}
    \centering
    \includegraphics[width=\linewidth]{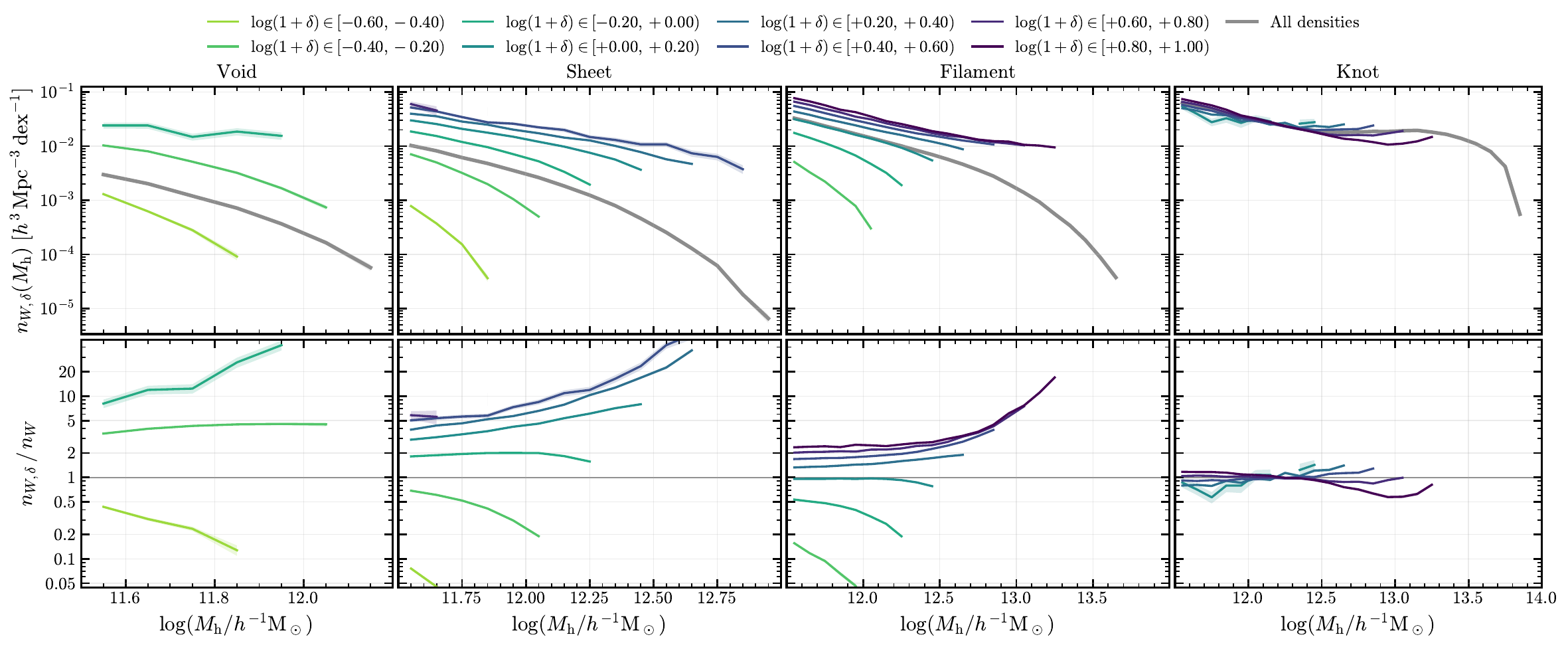}
    \caption{\textbf{Halo mass function at fixed web type, split by local density.}
    Columns show the four web environments: void, sheet, filament, and knot. Upper panels present the HMFs, $n_{W,\delta}(M_{\rm h})$ in different $W$ and $\delta$ (color-coded), together with $n_W(M_{\rm h})$ of the same web type without any density split (thick gray line). Lower panels give the ratio $n_{W,\delta}/n_W$. Shaded regions show the $1\sigma$ uncertainties, estimated by jackknife resampling of $N_{\rm jk}=4^3$ spatial subvolumes of the simulation box. Only the results that are calculated using $\ge30$ halos are drawn. We also exclude the results with $\log M_{\rm h}\ge 12.45+\log(1+\delta)$ (see text for the details).
    }
    \label{fig:hmf-web}
\end{figure*}

\section{Halo formation and Environment}
\label{sec:halo_env}

The cosmic-web classification introduced in Section~\ref{sec:simulation} and the local density are two different quantities to describe the environment. Here we measure halo properties as a joint function of both, and investigate how halos are correlated with the two environmental indicators. These can tell us how environments affect the halo formation.

\subsection{Environmental dependence of halo mass function}
\label{subsec:hmf}

We define the HMF within a given environment as,
\begin{equation}
    n_{W,\delta}(M_{\rm h}) = \frac{N(M_{\rm h},\delta,W)}{V(\delta,W)\,\Delta M_{\rm h}},
    \label{eq:hmf_wx}
\end{equation}
where $N(M_{\rm h},\delta,W)$ is the number of halos in the corresponding $(M_{\rm h},\delta,W)$ bin and $V(\delta,W)$ is the total volume of grid cells of web type $W$ and density $\delta$. Summing over the density bins gives $n_W(M_{\rm h})$, the HMF for web type $W$ as a whole; summing over web types instead gives $n_\delta(M_{\rm h})$, the HMF at fixed density. 
Figure~\ref{fig:hmf-web} presents these HMFs at fixed web type. For each web type, we also show the ratio between $n_{W,\delta}(M_{\rm h})$ and $n_W(M_{\rm h})$ in the bottom panels. 
We present results only for $\log M_{\rm h}< \log M_{\rm s}-0.5=12.45+\log(1+\delta)$, since at higher $M_{\rm h}$ the halo contribution to the density estimate becomes significant.
The results with less than $30$ halos are also excluded.

\begin{figure}
    \centering
    \includegraphics[width=\linewidth]{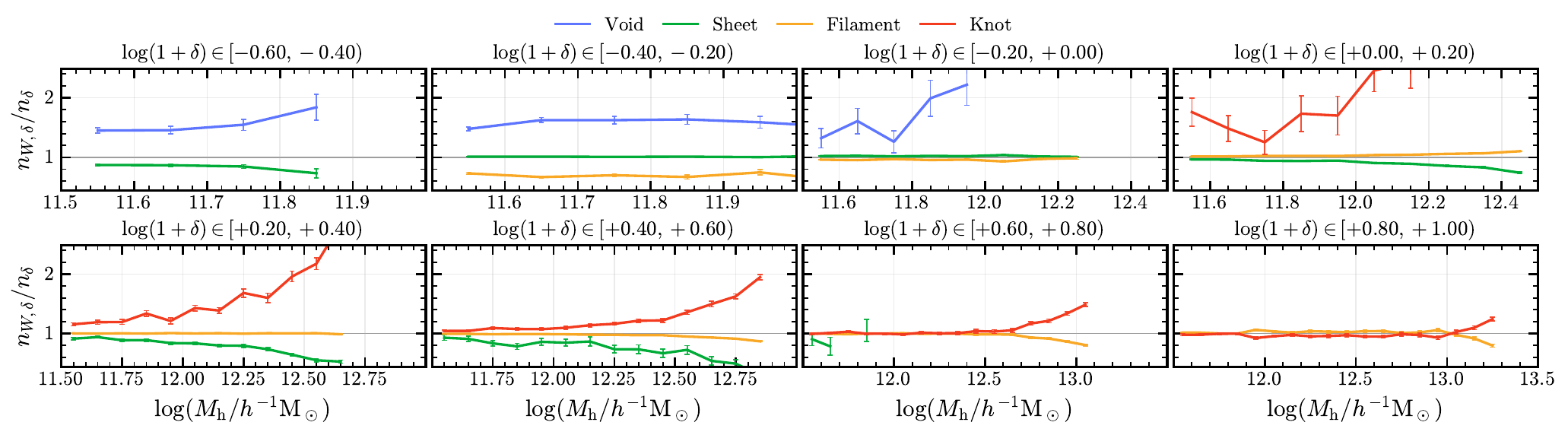}
    \caption{\textbf{The HMF ratio at fixed local density, split by web type.}
    Each panel corresponds to one $\delta$ bin, as labeled, and shows the HMF ratio $n_{W,\delta}/n_\delta$. Colors denote the web type. Error bars show the $1\sigma$ uncertainties, estimated by jackknife resampling of $N_{\rm jk}=4^3$ spatial subvolumes of the simulation box. Only web types that are sufficiently populated at a given density are drawn.}
    \label{fig:hmf-density}
\end{figure}

The gray curves ($n_W(M_{\rm h})$) reproduce the result of \citet{hahn2007properties}: the high-mass cut-off moves to larger masses from voids through sheets and filaments to knots, as expected from the conditional mass function of the excursion-set formalism \citep{bond1991excursion, bower1991evolution}. 
Within each web type, however, the HMF remains strongly ordered by local density. Across nearly the full range of $M_{\rm h}$ considered, the HMF in the void, sheet, and filament regions increases as $\delta$ increases. This makes sense because a denser environment supplies more material for halos to accumulate through accretion.

This rise is steep for $\delta<0$, but becomes more gradual for $\delta>0$. For example, in voids and sheets, as $\log(1+\delta$) increases from $\sim-0.5$ to $-0.1$, $n_{W,\delta}(M_{\rm h})$ at $\log M_{\rm h}/\msun \sim 11.8$ grows by a factor of 100 or more. 
By comparison, when $\log(1+\delta$) increases from $\sim0.1$ to $0.5$, the HMF in sheets shows only a modest change, increasing by a factor of $<2$.
One potential explanation is that dark matter in denser regions also has higher velocities \citep{wang2007environmental}, which in turn suppresses halo accretion.
The shape of the HMF also shows a mild dependence on $\delta$, with the slope flattening at higher $\delta$.  The variation of the slope with $\delta$ is also stronger in low $\delta$ environments and weaker in high $\delta$ environments.
In knot regions there is no clear dependence on $\delta$. This is in fact compatible with the behavior in the other web environments, because the $\delta$ in knots is typically greater than zero.

If density alone fully characterized the environment, then at fixed $\delta$ the HMF would not vary with web type. In Figure~\ref{fig:hmf-density}, we therefore contrast the HMFs of different web types while holding $\delta$ constant. Overall, we find a complicated but generally weak dependence on web type. For $\delta<0$, halos form most readily in voids and least readily in filaments. The HMF slopes are broadly comparable across void, sheet, and filament environments. A plausible reason is that halos in voids lack significant surrounding structures, so halo accretion is less influenced by the environment and thus can effectively accrete the surrounding material.  
For $\delta>0$, the trend changes. The halo number density is highest in knots and lowest in sheets. Moreover, the web type dependence is nearly negligible for low-mass halos but becomes more pronounced at the high-mass end. Since massive halos assemble primarily by accreting smaller halos, and knot regions contain more small halos than other web types (see next section), massive halos can build up more efficiently in knots.

We now compare our result with previous works.
\citet{alonso2015halo} reported that, at fixed density, the HMFs measured in different web environments are consistent with one another to within $\sim13$ percent when using smoothing scales of $4$ and $10\,h^{-1}{\rm Mpc}$. 
We use the same smoothing scales for both $\lambda_{\rm th}=0.0$ and the threshold adopted in their study, and find that the influence of web classification on the HMF weakens as $R_{\rm s}$ increases. For $R_{\rm s}=10\mpc$, the dependence on web type nearly vanishes once $\delta$ is fixed. Hence, the discrepancy between our results and those of \citet{alonso2015halo} is primarily attributed to the choice of smoothing scale. This indicates that the small-scale tidal field plays an important role in halo formation.
\citet{metuki2016abundance} reported instead that the HMF is governed by cosmic web type rather than by local density. In their analysis, they used different smoothing lengths to identify the cosmic web ($R_{\rm s}\sim1\mpc$) and to estimate the local density ($R_{\rm s}=4\mpc$). We suspect that this discrepancy is likely driven by the use of mismatched smoothing scales. 
To examine this, we run tests where web classification is performed with $R_{\rm s}=1\,h^{-1}{\rm Mpc}$ or $2\mpc$, while local density is computed with $R_{\rm s}=4\,h^{-1}{\rm Mpc}$ or $8\mpc$. Under this setup, relative to our fiducial results, the HMF shows a much stronger dependence on web type at fixed $\delta$, whereas at fixed web type its dependence on $\delta$ becomes much weaker and in some cases vanishes. This behavior is broadly in line with the trend presented in \citet{metuki2016abundance}. We therefore contend that applying different $R_{\rm s}$ values for the two environmental measures is not a fair comparison, since environmental effects depend sensitively on the smoothing scale.

Overall, the halo mass function is strongly dependent on local density and is relatively weakly but significantly related to cosmic web. Therefore, both the supply of material and the dynamical environment affect the formation of halos, consistent with the theoretical expectation \citep{sheth2001ellipsoidal}. We find that the environmental dependence changes dramatically around $\delta\sim 0$, providing an important clue for understanding halo formation.  

\subsection{Environmental dependence of formation redshift}
\label{subsec:formation_redshift}

\begin{figure}[ht!]
    \centering
    \includegraphics[width=\linewidth]{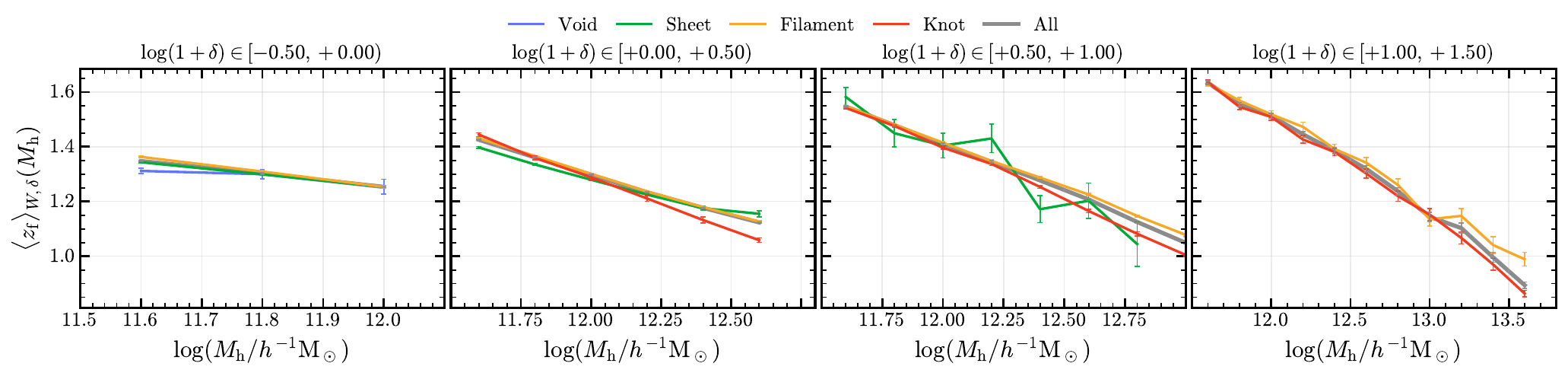}
    \caption{\textbf{Mean formation redshift at fixed local density, split by web type.}
    Each panel corresponds to one $\delta$ bin of width $0.5$~dex, as labeled. Colors denote the web type. Error bars show the $1\sigma$ uncertainties, estimated by jackknife resampling. }
    \label{fig:zf-density}
\end{figure}

\begin{figure*}
    \centering
    \includegraphics[width=\linewidth]{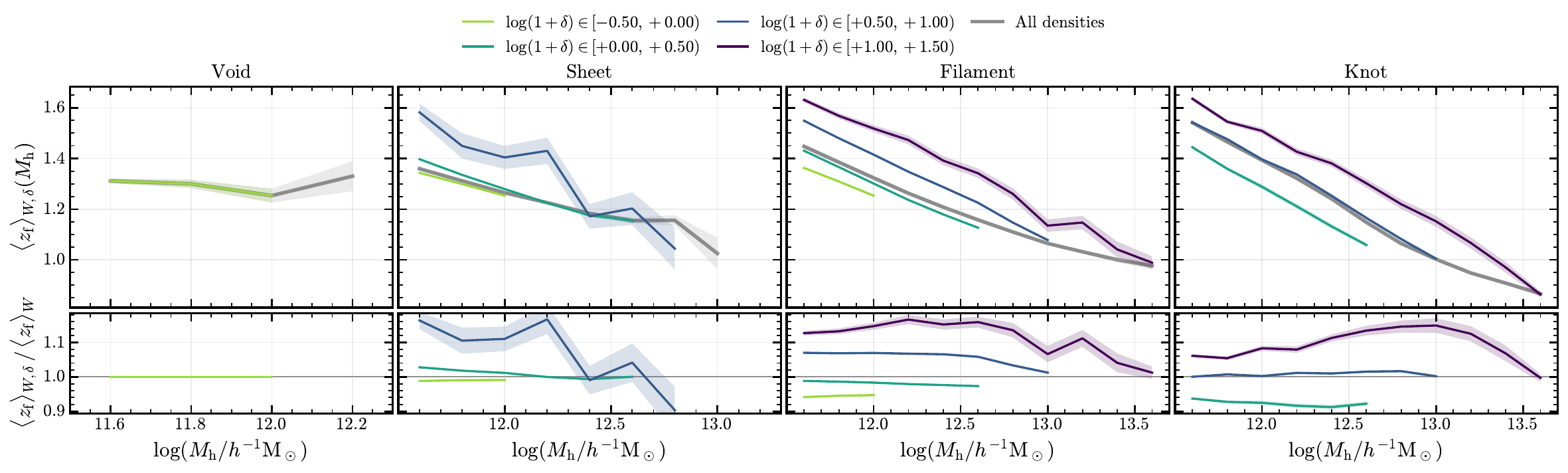}
    \caption{\textbf{Mean formation redshift at fixed web type, split by local density.}
    Layout follows Figure~\ref{fig:hmf-web}. Upper panels give the mean formation redshift $\langle z_{\rm f}\rangle_{W,\delta}(M_{\rm h})$ of halos (color-coded), together with $\langle z_{\rm f}\rangle_W(M_{\rm h})$ (thick gray line); lower panels show the ratio $\langle z_{\rm f}\rangle_{W,\delta}/\langle z_{\rm f}\rangle_W$. Shaded regions show the $1\sigma$ uncertainties, estimated by jackknife resampling. The size of $\delta$ bins is $0.5$~dex, wider than in Figure~\ref{fig:hmf-web}.}
    \label{fig:zf-web}
\end{figure*}

\citet{hahn2007properties} reported that, at low mass, halos are older in knots and filaments than in sheets and voids, by nearly $20$ percent between knots and voids, and the difference vanishes by $\log M_{\rm h}/\msun\simeq12.5$. It is also known that different web types exhibit very different density ranges. It is thus necessary to examine which environmental parameter dominates this environmental dependence.  We calculate the mean formation redshift ($\langle z_{\rm f}\rangle_{W,\delta}$) within each $(\delta,W)$ bin as functions of $M_{\rm h}$.  To reduce statistical uncertainties, we adopt a wider $\delta$ binning of 0.5 dex. We also note that we report results only for $\log M_{\rm h}< \log M_{\rm s}-0.5$.

In Figure \ref{fig:zf-density}, we compare $\langle z_{\rm f}\rangle$ across different web types within the same $\delta$ bin. As shown, for a fixed $\delta$ the differences related to the web type disappear largely, with only a faint residual trend visible at the high-mass end. Since a large $\delta$ bin size is adopted in the analysis, the residual trend might be caused by the density variation in each bin. Therefore, the environmental dependence of halo assembly history is mainly driven by the local density rather than the cosmic web.
As shown in Figure \ref{fig:zf-web}, within each web type, there is a significant dependence of $\langle z_{\rm f}\rangle$ on $\delta$. On average, halos residing in high-$\delta$ regions form earlier (are older) than those in low-$\delta$ regions. This behavior
is evident in sheet, filament, and knot environments. 

The trend appears to be broadly consistent with the assembly bias effect, in which older halos are more strongly clustered than younger halos of the same mass \citep[e.g.][]{gao2005age}. 
However, assembly bias is strong for low-mass halos and becomes nearly negligible at $\log M_{\rm h}/\msun\ge13$, whereas the trend shown here looks similar across the $M_{\rm h}$ range considered. 

\section{3D cosmic web cross-correlation function}
\label{sec:web_halo_correlation}

\subsection{Method}
\label{sec:method}

Here, we introduce the three-dimensional cosmic-web cross-correlation function (3D-WCCF). For a web cell $W$ and a halo $h$, we calculate their separation $\mathbf{r}=(r_1,r_2,r_3)$ as, 
\begin{equation}
    r_i(W,h) = \left| (\mathbf{x}_h-\mathbf{x}_W) \cdot\mathbf{e}_i(\mathbf{x}_W) \right|,\qquad i=1,2,3,
    \label{eq:folded_sep}
\end{equation}
where $\mathbf{x}_W$ and $\mathbf{x}_h$ are the positions of the web cell and the halo, respectively, and $\mathbf{e}_i(\mathbf{x}_W)$ is the $i$-th tidal eigenvector at $\mathbf{x}_W$. Taking the absolute value folds the statistic into the first octant and is required because the sign of a tidal eigenvector is arbitrary. Then the 3D-WCCF is defined via Peebles-Hauser (PH) estimator \citep{peebles1974statistical}
\begin{equation}
    {\xi}_{\rm 3D}(\mathbf{r})\equiv{\xi}_{\rm 3D}(r_1,r_2,r_3)=\frac{D_WD_h(\mathbf{r})}{R_WR_h(\mathbf{r})}-1.
    \label{eq:3d_wccf}
\end{equation}
The 3D-WCCF measures the distribution of halos around the cosmic web in the local frame of the tidal tensor.  It preserves anisotropic information about the surrounding large-scale structures. Although the definition is general to all types of tracers, the present implementation and analysis use halos only, which are the direct tracer of the dark matter density field and bridge between matter field and galaxies.

A conventional isotropic correlation function measures whether halos are preferentially found at a given distance from a cosmic-web environment. In this section, we use the isotropic halo--web cross-correlation function (hereafter called 1D-WCCF), $\xi_{\rm 1D}(r)$, as a baseline for the 3D-WCCF. The 1D-WCCF quantifies whether halos are over- or underabundant around a given web environment, while the 3D-WCCF further reveals how that abundance is distributed among the most-compressing ($\mathbf{e}_1$), intermediate ($\mathbf{e}_2$), and most-extending ($\mathbf{e}_3$) directions of the local tidal field.

Figure \ref{fig:main_xi} presents the 1D-WCCF and 3D-WCCF. For convenience, we denote $\xi_{\rm 3D}({\bf r})$ along ${\bf e}_i$ as $\xi_{\rm 3D}(r_i)$. For instance, $\xi_{\rm 3D}(r_1)$ refers to the correlation function as a function of $r_1$, while the other two radial components are held fixed at the innermost bins, i.e. $r_2\sim0$ and $r_3\sim0$.  We divided halos into six $M_{\rm h}$ bins as indicated in the panels.
In general, $\xi_{\rm 1D}(r)$ is positive and declines with radius for knots and filaments, whereas it is negative and rises with radius for sheets and voids. This agrees with the picture that knots and filaments trace high-density regions, while sheets and voids correspond to low-density environments.  
For all six $M_{\rm h}$ bins, we consistently find $\xi_{\rm 3D}(r_3)\ge\xi_{\rm 3D}(r_2)\ge\xi_{\rm 3D}(r_1)$. The 3D-WCCF also shows a number of notable characteristics that are not apparent in the 1D-WCCF, which we describe below.

\subsection{Anisotropy in halo distributions}
\label{subsec:main_directional}

\begin{figure*}[!t]
    \centering
    \includegraphics[width=\linewidth]{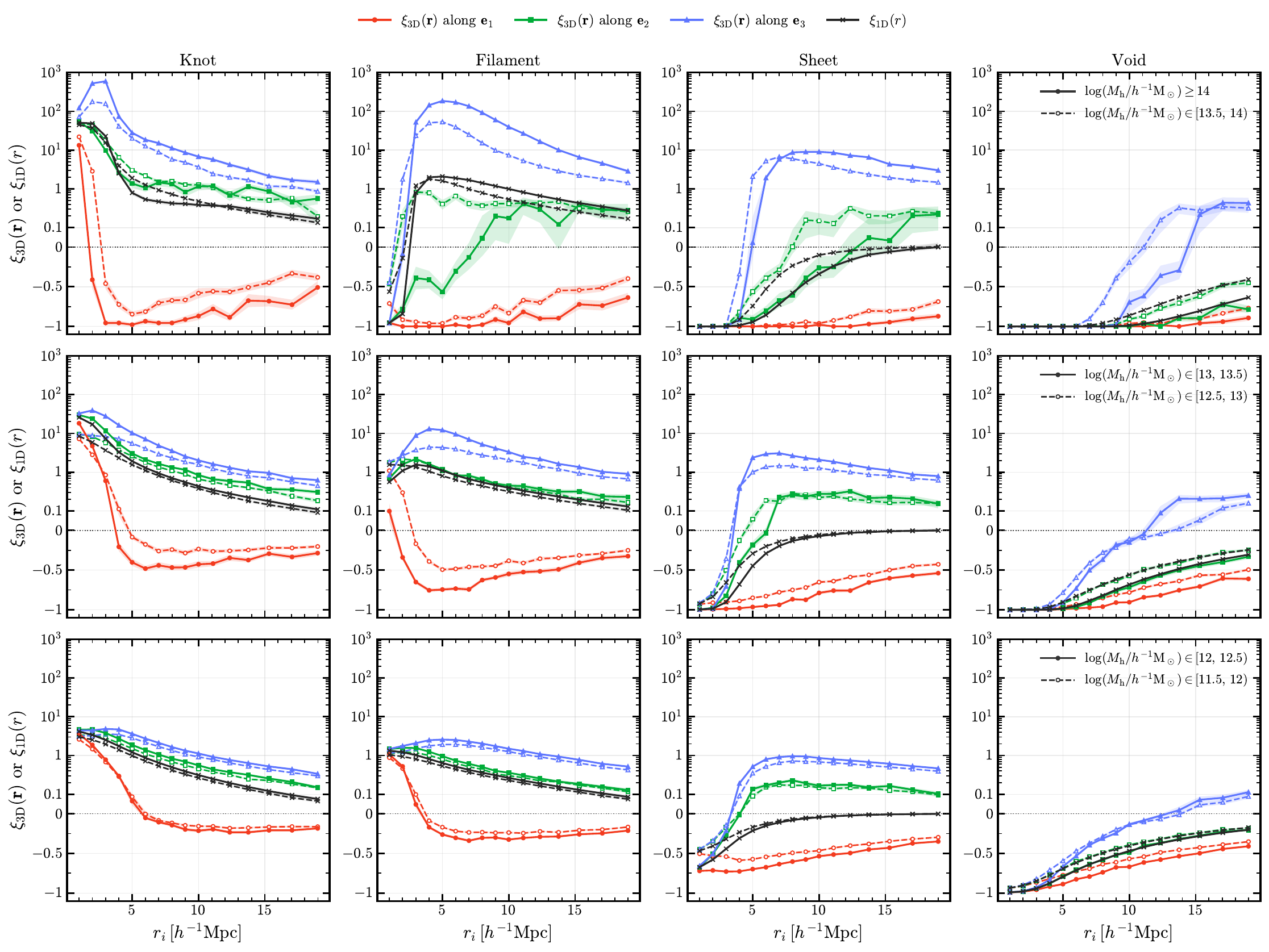}
    \caption{\textbf{Three-dimensional cross-correlation functions between cosmic-web and halos.}
    Columns correspond to knot, filament, sheet, and void, while each row shows the results for two halo-mass bins as indicated in the rightmost panels. Red, green and blue lines show the 3D-WCCF along ${\rm e}_1$($\xi_{\rm 3D}(r_1)$), along ${\rm e}_2$($\xi_{\rm 3D}(r_2)$) and along ${\rm e}_3$($\xi_{\rm 3D}(r_3)$), and black lines show the 1D-WCCF result $\xi_{\rm 1D}(r)$. See text for the definition of these quantities.}
    \label{fig:main_xi}
\end{figure*}

For knots, the 3D-WCCF along the three principal directions generally declines as the radius increases.
In the innermost radial bins, the correlations between massive halos ($\log M_{\rm h}/\msun\ge14$, hereafter referred to as clusters) and knots are close to 100, whereas the correlations with the other web types are all nearly $-1$. We note that $-1$ indicates that halos are not located in this region.
Therefore, nearly all clusters reside in knots. The radial decline is most pronounced along ${\bf e}_1$, the strongest-compression axis. For the knot–cluster correlation, 
$\xi_{\rm 3D}(r_1)$ rapidly drops to $\sim-1$ at about 2$\mpc$ and stays near $\sim-1$ or $<-0.5$ over the full range considered ($r<20\mpc$). By comparison, $\xi_{\rm 3D}(r_2)$ and $\xi_{\rm 3D}(r_3)$ decrease rapidly for $r<5\mpc$; afterward, $\xi_{\rm 3D}(r_2)$ remains approximately constant around $1$, while $\xi_{\rm 3D}(r_3)$ continues to decline gradually and reaches $1$ at $r\sim20\mpc$. Therefore, these results suggest that the clusters preferentially lie in the ${\bf e}_2-{\bf e}_3$ plane around knots. 
This may help explain why nearby galaxy clusters in the local Universe are found close to the supergalactic plane \citep{tully1986alignment, bohringer2021cosmic, peebles2022extended}. 
For halos with lower mass, the trends are similar, but become much weaker.

For filaments, halos are expected to preferentially lie along the ${\bf e}_3$ direction.
 This expectation agrees with our finding that $\xi_{\rm 3D}(r_3)$ is consistently larger than $\xi_{\rm 3D}(r_1)$ and $\xi_{\rm 3D}(r_2)$ at all separations and across all halo masses. At $r\sim10\mpc$, the ratio between $\xi_{\rm 3D}(r_3)$ and $\xi_{\rm 3D}(r_2)$ is $\sim 200$ for the clusters and 4 for the lowest-mass halos.
For halos with $\log M_{\rm h}/\msun<13.5$, $\xi_{\rm 3D}(r_1)$ decreases rapidly, changing from positive to negative at a radius ranging from 1.5$\mpc$ to 3.5$\mpc$, with the precise transition radius varying with the halo mass. This transition scale could be associated with the filament width \citep{zhang2024statistical}. 
In contrast, $\xi_{\rm 3D}(r_2)$ remains positive at all radii, implying that filaments have a substantial mass extension along ${\bf e}_2$. A plausible explanation is that, in the ${\bf e}_2$ direction, filaments are linked to other filaments and sheets.
Interestingly, for the clusters, both $\xi_{\rm 3D}(r_1)$ and $\xi_{\rm 3D}(r_2)$ become negative at $r_2<8\mpc$. This suggests that the clusters set the ${\bf e}_3$ direction of the filaments in their surroundings ($r<8\mpc$). Furthermore, $\xi_{\rm 3D}(r_1)$ for clusters stays near $-1$ at $r<15\mpc$, indicating that clusters can influence filament orientations even when they are located at large distances from the filaments.

The correlation functions for sheets differ from those of the two denser web types. In the innermost radius bins, $\xi_{\rm 3D}(r_i)$ is consistently negative, indicating that sheets are less likely to be associated with halos, in agreement with the HMF (Figure \ref{fig:hmf-web}).  
On larger scales, $\xi_{\rm 3D}(r_2)$ and $\xi_{\rm 3D}(r_3)$ rise rapidly and turn positive. The scale of this sign change strongly depends on $M_{\rm h}$ for $\xi_{\rm 3D}(r_2)$, while for $\xi_{\rm 3D}(r_1)$ it shows only a weak dependence on $M_{\rm h}$. 
Beyond the transition scale, $\xi_{\rm 3D}(r_2)$ is nearly independent of radius in all examined $M_{\rm h}$ bins, and $\xi_{\rm 3D}(r_3)$ exhibits only a mild dependence on radius. For halos with $\log M_{\rm h}/\msun<12.5$, on scales between $5\mpc$ and $15\mpc$, we typically find $\xi_{\rm 3D}(r_3)\sim0.7$, $\xi_{\rm 3D}(r_2)\sim0.2$, and $\xi_{\rm 3D}(r_1)\sim-0.5$. Overall, this approximately corresponds to a pancake-like configuration: a structure with density slightly above the mean, lying in the ${\bf e}_2-{\bf e}_3$ plane, showing weak radial variation but a pronounced dependence on orientation.

For voids, all correlation functions increase with increasing radius. Within $\sim5\mpc$, $\xi_{\rm 3D}({\bf r})$ is nearly isotropic and close to $-1$. However, at larger scales, the correlation along ${\bf e}_3$ becomes significantly larger than those along the other two axes. The difference between $\xi_{\rm 3D}(r_1)$ and $\xi_{\rm 3D}(r_2)$ is relatively small. Therefore, voids are on average an oblate structure, compressed along ${\bf e}_3$ direction. In the scale range considered, $\xi_{\rm 3D}(r_1)$ and $\xi_{\rm 3D}(r_2)$ are always negative, while $\xi_{\rm 3D}(r_3)$ becomes positive at larger radius.
The transition radius for low-mass halos ($\log M_{\rm h}/\msun<12.5$) is nearly 13$\mpc$. Therefore, the typical radius of voids is around 13$\mpc$. This radius is close to the mean effective radius of voids identified using different approaches\citep{zhang2026robustness}. However, it is worthwhile to note that the void sizes measured along the other principal axes of the tidal tensor are much longer, $>20\mpc$, as demonstrated in Figure \ref{fig:main_xi}.

\subsection{Anisotropy of assembly bias}
\label{sec:assembly_bias}

\begin{figure*}[!t]
    \centering
    \includegraphics[width=\linewidth]{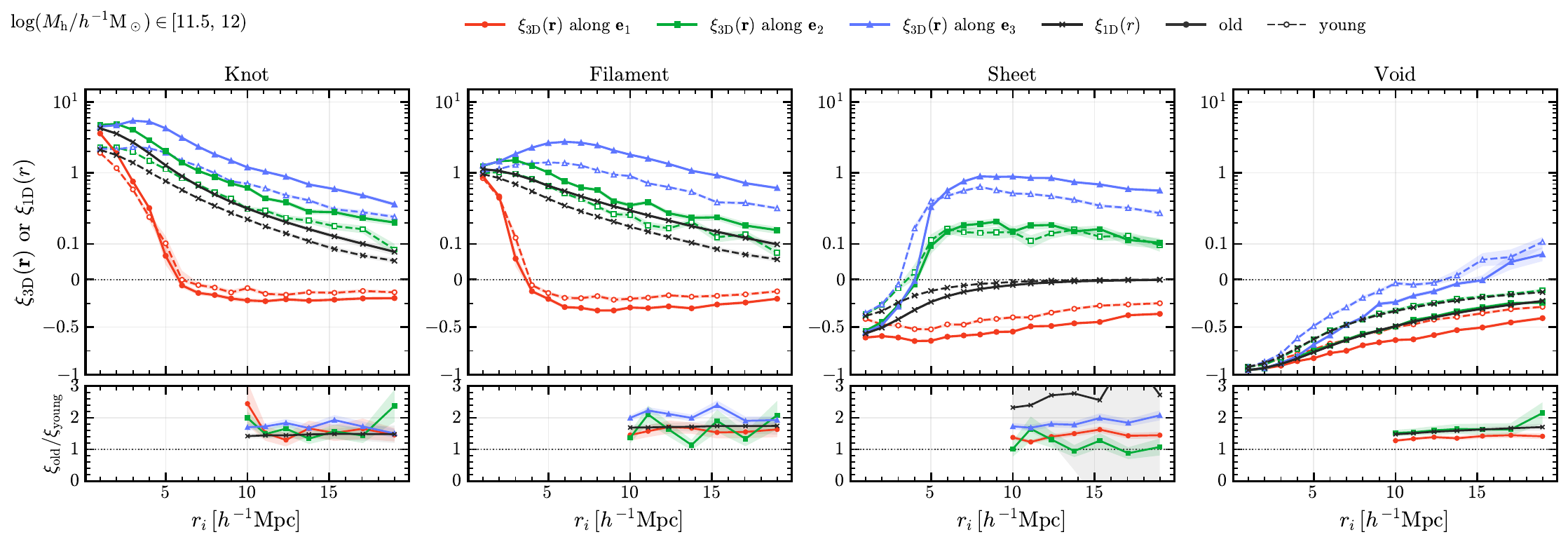}
    \caption{\textbf{Three-dimensional cosmic-web cross-correlation function for old and young halo samples.} Here we only show the results for the lowest mass bin.
    In upper panels, layout follows Figure ~\ref{fig:main_xi}. The solid lines show the old halos and dashed lines show the young halos. Lower panels present the ratio $\xi_{\rm old}/\xi_{\rm young}$, here we use $\xi$ to denote $\xi_{\rm 3D}$ or $\xi_{\rm 1D}$. In these panels we cut the scale at $r_i = 10\, h^{-1}{\rm Mpc}$ to avoid negative ratio. The result along ${\rm e}_3$ for the void is also not presented because $\xi_{\rm 3D}(r_3)$ changes its sign at $r>10\mpc$. }
    \label{fig:main_xi_by_age}
\end{figure*}

It is therefore worthwhile to investigate whether the assembly bias effect is anisotropic. To this end, within each mass bin, we further divide halos into two subsamples, “old” and “young”, defined respectively as the upper and lower $20\%$ tails of the $z_{\rm f}$ distribution in that bin. To eliminate any remaining mass dependence inside a given mass bin, we apply weights to halos in both subsamples so that each reproduces the mass distribution of the full halo population in that bin. 

Figure ~\ref{fig:main_xi_by_age} presents the 3D-WCCF and the 1D-WCCF for young and old samples.
As shown, both young and old halos satisfy $\xi_{\rm 3D}(r_3)\ge\xi_{\rm 3D}(r_2)\ge\xi_{\rm 3D}(r_1)$ in a consistent manner. Moreover, the form of the correlation functions does not depend on the halo age. In particular, for high-mass halos, $\xi_{\rm 3D}({\bf r})$ is indistinguishable between the two subsamples. 
This is in line with earlier work indicating that the assembly-bias effect associated with $z_{\rm f}$ becomes negligible for massive halos. We therefore show only the results for the lowest-mass halos.

As can be seen, when the correlation function is positive, old halos tend to have a stronger correlation function than young halos. When the correlation function is negative, old halos tend to exhibit a weaker correlation with cosmic web. This trend can be seen in both 3D-WCCF and 1D-WCCF. Apparently, the spatial distribution of old halos shows stronger anisotropy than that of young halos. As shown above, older halos tend to form in the denser region. Therefore, old halos are preferentially associated with filaments and knots, and thus exhibit stronger anisotropy in the cosmic web correlation functions. 

To characterize the anisotropy of the assembly bias, we calculate the relative bias, $b_{\rm r}$, which is defined as the ratio of $\xi_{\rm 3D}({\bf r})$ of old and young halos. Since the relative bias is roughly constant at $r\ge10\mpc$, we only present the bias at this scale, which is also where the linear bias is measured. 
The relative bias is almost always larger than 1 across different web types and directions, indicating a significant assembly bias signal. For knots, the relative bias along the three axes is similar, suggesting that the assembly bias is nearly isotropic with respect to knots. The anisotropy is significant in filaments, sheets and voids. 
In general, the relative bias along ${\bf e}_3$ is stronger than along the other two axes.

Our results agree with the prediction of \citet{musso2018cosmic} that early-forming halos lie preferentially along the filament axis.
More recently, \citet{ma2026anisotropic}, using each halo's own major axis as the reference direction, find the anisotropic assembly bias to be weak for formation time and strong for spin and shape. The major axis of a halo aligns only partially with the ${\bf e}_3$ axis of the local tidal field. The two results are therefore compatible, and together they suggest that the age dependence of clustering is organized by the geometry of the environment rather than by the orientation of the halo.

\section{Summary}
\label{sec:summary}

Galaxies and their host dark matter halos form and grow within the cosmic web. In this paper, we have used the ELUCID simulation at $z=0$ to study how the local environment affects the formation of dark matter halos and their spatial distribution around the cosmic web. We first separated the effects of the local density and the cosmic-web type, derived respectively from the trace and the eigenvalues of the tidal tensor, on the halo mass function and on the mean formation redshift, restricting the analysis to halos at least $0.5$~dex below the mass $M_{\rm s}$ contained in the smoothing kernel, so that the measured density reflects the environment rather than the halo itself. We then introduced the three-dimensional cosmic-web cross-correlation function (3D-WCCF), in which the separation between a web cell and a halo is expressed in the local tidal eigenframe, and used it to measure the anisotropic distribution of halos around each web type and its dependence on halo formation time. Halos with a backsplash history are excluded throughout.
Our main conclusions are as follows.

At fixed web type, the HMF in voids, sheets, and filaments increases strongly with the local density. The increase is steep for $\delta<0$ and becomes gradual for $\delta>0$, and the slope of the HMF flattens with increasing density. In knots, where $\delta$ is typically positive, no clear density dependence is found. At fixed density, the dependence on web type is weak but significant: for $\delta<0$ halos form most readily in voids and least readily in filaments, whereas for $\delta>0$ the HMF is highest in knots and lowest in sheets, with the difference growing toward high mass. Therefore, both the supply of material and the dynamical environment affect the formation of halos, with the environmental dependencies changing or reversing around $\delta\sim0$.  The apparently contradictory conclusions of \citet{alonso2015halo} and \citet{metuki2016abundance} can both be traced to the choice of smoothing scale: the web-type dependence at fixed density weakens as $R_{\rm s}$ increases and nearly vanishes at $R_{\rm s}=10\,h^{-1}{\rm Mpc}$, while measuring the density on larger scales than web type strengthens the dependence of web types and suppresses the dependence of density.
 
At fixed web type, halos in denser regions form earlier, over the whole mass range considered. At fixed density, the differences between web types largely disappear, leaving only a faint residual trend at the high-mass end that may reflect the density variation within the wide density bins. The environmental dependence of halo assembly is therefore driven mainly by the local density, and the dependence on web type reported by \citet{hahn2007properties} can be attributed largely to the different densities of the web types. Since backsplash halos are excluded, the earlier formation in dense regions is more likely due to tidal truncation or heating than to strong tidal stripping \citep{wang2007environmental, hahn2009tidal, mansfield2020three}. 

The 3D-WCCF provides rich anisotropic information on the halo distribution. 
Around all four web types and for all halo masses, $\xi_{\rm 3D}(r_3)\ge\xi_{\rm 3D}(r_2)\ge\xi_{\rm 3D}(r_1)$: halos are distributed preferentially along the axis of weakest compression and avoid the axis of strongest compression. Nearly all halos with $M_{\rm h}\ge10^{14}\,h^{-1}{\rm M_\odot}$ reside in knots and lie preferentially in the $\mathbf{e}_2$--$\mathbf{e}_3$ plane around them, which may help explain the supergalactic plane \citep{tully1986alignment, bohringer2021cosmic, peebles2022extended}, and they also set the orientation of the surrounding filaments. Around filaments, 
the correlation along ${\bf e}_3$ is much higher than along the other axes, while the correlation of halos below $10^{13.5}\,h^{-1}{\rm M_\odot}$ along $\mathbf{e}_1$ changes sign at $1.5$--$3.5\,h^{-1}{\rm Mpc}$, a scale that may be associated with the filament width. Around sheets, the halo distribution corresponds to a pancake-like structure in the $\mathbf{e}_2$--$\mathbf{e}_3$ plane. Around voids, the halo deficit ends at $\sim13\,h^{-1}{\rm Mpc}$ along $\mathbf{e}_3$, close to the typical void radius found with other methods, but extends beyond $20\,h^{-1}{\rm Mpc}$ along the other two axes. Apparently, the 3D-WCCF provides important information about the geometry and size of these cosmic-web structures.

At fixed mass, old halos are more strongly correlated with the cosmic web than young ones, with a relative bias $b_{\rm r}>1$ for almost all web types and directions, while for massive halos the two subsamples are indistinguishable. The assembly bias is nearly isotropic around knots but anisotropic around filaments, sheets, and voids, where it is generally strongest along $\mathbf{e}_3$. This agrees with the prediction of \citet{musso2018cosmic} that early-forming halos lie preferentially along the filament axis. 

In future work, we plan to extend the 3D-WCCF to other tracers, such as the matter field and galaxies, and perform it on observations. For galaxy surveys such as SDSS, the ELUCID reconstructed density field can likewise be used to probe the relation between the galaxy distribution and the tidal field  \citep{wang2016elucid, zhang2025unexpected}. By comparing the 3D-WCCF of halos, simulated galaxies and observed galaxies, we can probe the baryonic physics process and observed large-scale structure. Measuring the web--matter cross-correlation $\xi_{\rm Wm}$ in particular would allow the absolute, direction-dependent halo bias rather than only the relative bias between two halo subsamples. 

To sum up, the halo--web relation is richer than either of the scalars conventionally used to describe environment. The local density and the web class each carry information the other does not, and their relative importance depends on the halo property; beyond both lies the directional structure of the tidal field, which the 3D-WCCF recovers and which isotropic statistics remove by construction. Modeling these relations will be important for assessing tidal-dependent systematics in bias, intrinsic-alignment, and redshift-space-distortion analyses, and for incorporating halo secondary properties into galaxy--halo models.

The derived data catalogs and code for analysis are available upon request to the corresponding authors.

\section*{Acknowledgements}
 
This work is supported by the National Natural Science Foundation of China (NSFC, Nos. 12595312, 12192224 ). HYW thanks the support of CAS Project for Young Scientists in Basic Research, Grant No. YSBR-062, the New Cornerstone Science Foundation through the XPLORER PRIZE, and the China Manned Space Program
with grant no. CMS-CSST-2025-A04. 
The work is supported by the Supercomputer Center of University of Science and Technology of China.

\bibliography{ref}{}
\bibliographystyle{aasjournalv7}

\end{document}